\documentclass[twocolumn]{webofc}

\usepackage[varg]{txfonts}   
\begin{document}

\def\bb    #1{\hbox{\boldmath${#1}$}}
 \def\oo    #1{{#1}_0 \!\!\!\!\!{}^{{}^{\circ}}~}  
 \def\op    #1{{#1}_0 \!\!\!\!\!{}^{{}^{{}^{\circ}}}~}
\def\blambda{{\hbox{\boldmath $\lambda$}}} 
\def\eeta{{\hbox{\boldmath $\eta$}}}
\def\bxi{{\hbox{\boldmath $\xi$}}} 
\def\bzeta{{\hbox{\boldmath $\zeta$}}}
\def\sD{D \!\!\!\!/}
\def\sd{\partial \!\!\!\!/}
\def\qcdu{{{}_{ \rm QCD}}}   
\def\qedu{{{}_{\rm QED}}}   
\def\qcdd{{{}^{ \rm QCD}}}   
\def\qedd{{{}^{\rm QED}}}   
\def\qcd{{{\rm QCD}}}   
\def\qed{{{\rm QED}}}   
\def\2d{{{}_{\rm 2D}}}         
\def\4d{{{}_{\rm 4D}}}         

%
\title{QED Mesons, the QED Neutron,
and the Dark Matter}
%
%

\author{\firstname{Cheuk-Yin} \lastname{Wong}\inst{1}\thanks{ 
Email:wongc@ornl.gov}}

\institute{Physics Division, Oak Ridge National Laboratory\!\thanks{
    This manuscript has been authored in part by UT-Battelle, LLC,
    under contract DE-AC05-00OR22725 with the US Department of Energy
    (DOE). The US government retains and the publisher, by accepting
    the article for publication, acknowledges that the US government
    retains a nonexclusive, paid-up, irrevocable, worldwide license to
    publish or reproduce the published form of this manuscript, or
    allow others to do so, for US government purposes. DOE will
    provide public access to these results of federally sponsored
    research in accordance with the DOE Public Access Plan
    (http://energy.gov/downloads/doe-public-access-plan), Oak Ridge,
    Tennessee 37831, USA }\!\!\!, ~\,Oak Ridge, 
    Tennessee 37831, USA }

\abstract{
Schwinger's boson solution for massless fermions in QED in 1+1D has
been applied and generalized to quarks interacting in QED and QCD
interactions, leading to stable and confined open-string QED and QCD
boson excitations of the quark-QCD-QED system in 1+1D.  Just as the
open-string QCD  excitations in 1+1D can be the idealization of
QCD mesons with a flux tube in 3+1D, so the open-string QED excitations
 in 1+1D may likewise be the idealization of QED mesons with
masses in the tens of MeV region, corresponding possibly to the
anomalous X17 and E38 particles observed recently.  A further search
for bound states of quarks interacting in the QED interaction alone
leads to the examination on the stability of the QED neutron,
consisting of two $d$ quarks and one $u$ quark.  Theoretically, the
QED neutron has been found to be stable and estimated to have a mass
of 44.5 MeV, whereas the analogous QED proton is unstable, leading to
a long-lived QED neutron that may be a good candidate for the dark
matter.
}

\maketitle

\section{Introduction}
\label{intro}

The quark-QCD-QED vacuum is the lowest-energy state of the
quark-QCD-QED system with quarks filling up the (hidden)
negative-energy Dirac sea and interacting with the QCD (quantum
chromodynamical) interaction and the QED (quantum electrodynamical)
interaction.  It is defined as the state with no valence quark above
the Dirac sea and no valence antiquark as  a hole below the Dirac sea.
A local disturbance will generate stable collective excitations of the
quark-QCD-QED system as in a fluid.  There are stable and confined
collective QCD excitations showing up as QCD mesons, such as $\pi$,
$\eta$, and $\eta'$. The question of our central interest is whether
there may also exist stable QED excitations of the quark-QCD-QED
system showing up as QED mesons in the mass region of many tens of
MeV, together with the related QED neutron with $d$-$u$-$d$ quarks
stabilized by the QED interaction.

Such a question may appear preposterous as one can surmise from the
following debate between a Wise Guy and an Explorer.  The Wise Guy
objects and argues that quarks interact in QCD and QED simultaneously.
When quarks and antiquarks interact, their color charges and electric
charges interact simultaneously. A stable QED excitation of the quarks
cannot be independent and cannot occur without the simultaneous QCD
excitation.  To these objections, the Explorer defends and replies
that the quark currents, as well as the QCD and the QED gauge fields,
are not single-element quantities.  They are 3$\times$3 color matrices
with 9 matrix elements that separate naturally into\break $\bb 3$
$\otimes$ $\bb 3^*$=$\bb 1$ $\oplus$ $\bb 8$, with a color-singlet
subgroup~$\bb 1$ and a color-octet subgroup $\bb 8$ executing
collective dynamics separately.  In the space-time arena, there can be
localized QCD excitations of the color-octet current leading to QCD
mesons. There can also be independent QED excitations of the
color-singlet current showing up as QED mesons at different energies.
The Wise Guy objects further and says that only the non-Abelian QCD
interaction can confine quarks, and the QED interaction does not
confine particles such as electrons and positrons.  Thereupon, the
Explorer explains that light quarks can be considered approximately
massless, and the application of the Schwinger's solution
\cite{Sch62,Sch63} indicates that massless quarks in QED are confined
in 1+1D.  The confined open string in 1+1D can be considered to be the
idealization of a flux tube in 3+1D.  Gribov showed that massless
fermions in QED in 3+1D are confined \cite{Gri82,Gri99}.  Furthermore,
lattice gauge calculations show that fermions in QED in 3+1D can be
confined under appropriate conditions \cite{Wil74,Mag20}.  Clearly,
whatever the outcome of the debate may be, it is necessary to confront
the predictions on the QED mesons and the QED neutron of the
quark-QCD-QED system with experiments.  The present investigation
facilitates such a confrontation.

\section{Generalization of Schwinger's  solution  from QED to (QED+QCD) in 1+1D}
\label{sec-1}
We first summarize the dynamics of the Schwinger's QED solution for
massless fermions in 1+1D as reviewed in \cite{Won94}.  Subject to a
QED gauge field disturbance $A^\mu$ with a coupling constant $g_\2d$
in 1+1D, the massless fermion field $\psi(x)$ satisfies the Dirac
equation,
\begin{eqnarray}
\gamma_\mu ( p^\mu - g_{\2d} A^\mu) \psi =0.
\end{eqnarray}
The disturbance $A^\mu$ instructs the fermion field $\psi$ how to
move, and 
through the Maxwell
equation,
\begin{eqnarray}
\hspace*{-0.3cm}\partial_\mu F^{\mu \nu}=\partial_\mu (\partial^\mu A^\nu -  \partial^\nu A^\mu)=g_\2d j^\nu = g_\2d \bar \psi \gamma^\nu \psi,
\end{eqnarray}
the fermion field $\psi$ in turn generates the current $j^\mu$ which
instructs the gauge field $A^\mu$ how to act.  A stable collective
excitation of the fermion-QED system occurs, when the disturbance
$A^\mu$ gives rise to the current $j^\mu$ which in turn leads to the
gauge field $A^\mu$ self-consistently.  By imposing the Schwinger
modification factor to ensure the gauge invariance of the fermion
Green's function, the fermion current $j^\mu(x) $ at the space-time
point $x$ induced by $A^\mu$ can be evaluated. After the singularities
from the left and from the right cancel, the fermion current $ j^\mu$
is found to  relate explicitly to the perturbing QED gauge field
$A^\mu$ by \cite{Sch62,Sch63}
\begin{eqnarray}
j^\mu = -\frac{g_\2d^2}{\pi }\left ( A^\mu - \partial ^\mu \frac{1}{\partial _\lambda \partial ^\lambda} \partial _\mu A ^\mu \right ) .
\end{eqnarray}
Upon substituting such a relation to the Maxwell equation (2), we get
$j^\mu$ and $A^\mu$ both satisfy the Klein-Gordon equation
\begin{eqnarray}
\hspace*{-0.8cm}\partial _\nu \partial ^\nu A^\mu + \frac{g_\2d^2}{\pi }A^\mu =0,  ~~{\rm and~} ~~
\partial _\nu \partial ^\nu j^\mu +\frac{g_\2d^2}{\pi }j^\mu =0,\!
\end{eqnarray}
 for a boson with a mass given by $m=g_\2d/\sqrt{\pi}$.

The Schwinger model can be generalized from QED to (QED+QCD)
\cite{Won10,Won20}.  Because of the three-color nature of quarks, the
quark currents and the QED and QCD gauge fields are 3$\times$3 color
matrices which can be expanded in terms of the nine generators of the
U(3) group,
\begin{eqnarray}
\hspace*{-0.6cm}j^\mu = \sum_{i=0}^8 j^\mu_i t^i,~~A^\mu = \sum_{i=0}^8 A^\mu_i t^i,~~~
t^0=\frac{1}{\sqrt{6} }
\begin{pmatrix}
1 & 0 & 0 \cr
0 & 1 & 0 \cr
0 & 0 & 1 \cr
\end{pmatrix},
\end{eqnarray}
where $t^0$ is the generator of the U(1) color-singlet subgroup and
$t^1,t^2,...,t^8$ are the eight generators of the SU(3) color-octet
subgroup.  Because the current and the gauge field of each subgroup
depend on each other within the subgroup, but do not depend on the
current and the gauge field of the other subgroup, the two different
currents and the gauge fields in their respective subgroups possess
independent stable QED and QCD collective excitations at different
energies.  For the QCD interaction, stable QCD collective excitations
can be attained in restricted variations which keep the orientation of
the unit vector $\tau^1$ in the color-octet generator space fixed
while their amplitudes $A_1^\mu$ are allowed to vary.  That is, with
\begin{eqnarray}
j^\mu =j_0^\mu \tau^0+ j_1^\mu \tau^1,~~~~~~A^\mu =A_0^\mu \tau^0+ A_1^\mu \tau^1,~~~~~~
\label{eq7}
\end{eqnarray}
 where $\tau^0$=$t^0$, $\tau^1$=$\sum_{i=1}^8 n_i t^i$, with fixed
 direction cosines $n_i$=2${\rm tr}\{\tau^1 t^i\}$ in the color-octet
 generator space.  With the above currents and gauge fields (6) in the
 color-singlet and the color-octet subgroups, Schwinger's solution for
 QED can be trivially generalized to (QED+QCD) leading to the currents
 and gauge fields satisfying their corresponding Klein-Gorden
 equations with masses depending on their respective coupling
 constants \cite{Won10,Won20}.
 
 Consequently, there are independent collective QED and QCD
 excitations of the quark-QCD-QED system in 1+1D where these
 excitations can be described as open-string states of $q\bar q$
 pairs.  We note perceptively that in 1+1D, the open string does not
 have a structure, but the coupling constant $g_\2d$ has the dimension
 of a mass. In contrast in 3+1D, the flux tube has a structure with a
 radius $R_T$, but the coupling constant $g_\4d$ is dimensionless.
 The 1+1D open string can be considered an idealization of the
 physical meson in 3+1D with a flux tube of radius $R_T$, if the
 coupling constants $g_\2d$ and $g_\4d$ in the two different
 space-time dimensions are related by \cite{Won10,Won20}
 \begin{eqnarray}
g_\2d^2=g_\4d^2/(\pi R_T^2)=4\alpha/R_T^2.
\end{eqnarray}
  Consequently, the masses of the QED and QCD mesons in 3+1D in the
  open-string description are approximately
\begin{eqnarray}
m_\qcd^2=4 \alpha_\qcd/\pi R_T^2,~~~~~~~ m_\qed^2=4 \alpha_\qed/\pi R_T^2.
\end{eqnarray}

\section{QCD and QED mesons as open strings}
To get a better determination of the QCD and QED meson masses, it is
necessary to take into account the flavor mixtures $D_{ij}$ and the
quark rest masses $m_f$.  Using the method of bosonization, we obtain
the semi-empirical mass formula for the neutral QCD mesons with
$N_f=3$ \cite{Won20}
\begin{eqnarray}
\hspace*{-0.8cm}m_i^2=\biggl  (\sum_{f=1}^{N_f} D_{if}\biggr  )^2 \,\frac{4\alpha_\qcd}{ \pi R_T^2} +
m_\pi^2 \sum_{f=1}^{N_f} \frac{m_f}{(m_u+m_d)/2} (D_{if})^2,
\label{qcd}
\end{eqnarray}
and for QED with $N_f=2$ and isospin $(I, I_3=0)$, 
 \begin{eqnarray}
m_I^2=\biggl [ \frac{Q_u+(-1)^IQ_d}{\sqrt{2}} \biggr ]^2\,
\frac{4\alpha_{{}_{\rm QED}}}{ \pi R_T^2} + m_\pi^2\frac{\alpha_{{}_{\rm QED}}}{\alpha_\qcd} .
\label{qed}
\end{eqnarray}
where $Q_u$=$2/3$ and $Q_d$=$-1/3$ are the charge numbers of $u$ and
$d$ quarks respectively.  With the flavor mixture and quark masses as
given by the Particle Data Group \cite{PDG19}, $\alpha_{{}_{\rm
    QED}}$=1/137, $\alpha_\qcd$=0.68$\pm$0.08, and $R_T$=0.40$\pm$0.04
fm, we obtain the theoretical QCD and QED meson masses \,in\,Table
\ref{tb1}.

\vspace*{-0.1cm}
\begin{table}[h]
\centering
\caption{The experimental and theoretical masses of neutral, $I_3$=0,
  QCD and QED mesons, obtained with the semi-empirical mass formula
  (\ref{qcd}) for QCD mesons and (\ref{qed}) for QED mesons. }
  \vspace{-0.2cm}\hspace*{-0.3cm}
\begin{tabular}{|c|c|c|c|c|c|}

\cline{3-5} \multicolumn{2}{c|}{}& &Experimental&Semi-empirical\\ 
\multicolumn{2}{c|}{}&$[I(J^\pi$)] & mass & mass
 \\ 
 \multicolumn{2}{c|}{}& & & formula \\ 
 \multicolumn{2}{c|}{}& & (MeV) & (MeV)  \\ 
\hline
QCD&$\pi^0$ &[1(0$^-$)] &\!\!134.9768$\pm$0.0005\!\!& 134.9$^\ddagger$  \\ 
\!\!meson\!\!& $\eta$ &[0(0$^-$)] &\!\!547.862$\pm$0.017\!\!&498.4$\pm$39.8~~  \\ 
& $\eta'$  &[0(0$^-$)] & 957.78$\pm$0.06& 948.2$\pm$99.6~\\ \hline
QED& X17&[0(0$^-$)] &\!\!16.94$\pm$0.24$^\#$& 17.9$\pm$1.5  \\ 
\!\!meson\!\!&E38&[1(0$^-$)] &37.38$\pm$0.71$^\oplus$ & 36.4$\pm$3.8\\ 
\hline
\end{tabular}
\vspace*{0.1cm}
\hspace*{-2.95cm}$^\ddagger$ Calibration mass~~~~~~~~~~~~~~~~~~~~~~~\\
\vspace*{-0.1cm}\hspace*{-0.90cm}$^\#$A. Krasznahorkay $et~al.$, arxiv:2104.10075 ~~~~\\
\hspace*{0.30cm}$^\oplus$\,K. Abraamyan $et~al.$, EPJ Web Conf       204,08004(2019)\\
\label{tb1}
\end{table}

\vspace*{-0.2cm}
We find from Table 1 that the open-string description
of the QCD and QED mesons is a reasonable concept and the anomalous
X17 \cite{Kra16} and E38 \cite{Abr19} observed recently may be QED
mesons.  The parent particles of the anomalous soft photons \cite{DEL10} may be QED mesons.  

\section{ Schwinger's QED boson as a relativistic two-body problem
  in 1+1 dimensions}
  
 It is desirable to construct a phenomenological two-body model for
 Schwinger's QED bound state in 1+1D involving a massless fermion and
 an antifermion interacting in an effective two-body QED interaction
 $\Phi_{12}$.  The interaction $\Phi_{12}$ must be calibrated to
 yield Schwinger's exact solution in field theory.  Neglecting spins,
 the relativistic two-body wave equations for the wave function $\Psi$
 for the QED interactions in 1+1 dimensions consist of two mass-shell
 constraints on each of  the interacting particles \cite{Cra83,Saz87},
\begin{subequations} 
\begin{eqnarray}
  {\cal H}_{1}|\Psi\rangle =\biggl  \{ p_{1}^{2}-m_{1}^{2}-\Phi _{12}(x_{12\perp})\biggr \} |\Psi\rangle =0,
\label{3p1a}
\\
{\cal H}_{2}|\Psi\rangle =\biggl  \{  p_{2}^{2}-m_{2}^{2}-\Phi_{12} (x_{21\perp}) \biggr \} |\Psi\rangle=0.
\label{3p1b}
\end{eqnarray}
\end{subequations} 
In the center-of-mass frame where the particle momentum is
$p_i=(\epsilon_i,q_i)$, the above two-body wave equations become
\begin{subequations} 
\begin{eqnarray}
\epsilon_1^2 |\Psi\rangle = \biggl\{ q_1^2 + m_1^2 +\Phi_{12}(x_{12\perp})\biggr\} |\Psi\rangle =0,
\\
\epsilon_2^2 |\Psi\rangle = \biggl\{ q_2^2 + m_2^2 +\Phi_{12}(x_{12\perp})\biggr\} |\Psi\rangle =0,
\end{eqnarray}
\end{subequations}
and the mass of the bound states is $m= \epsilon_1 +\epsilon_ 2$.  By
requiring $\Phi_{12}$ to give the mass $m$ = $g_\2d/\sqrt{\pi}$ in a
linearly confining potential to match Schwinger's exact solution for
massless fermions in QED in 1+1D, we find \cite{Won20a}
\begin{eqnarray}
\hspace*{-0.6cm}\Phi_{12}(x_{12\perp} )=\frac{2\epsilon_1 \epsilon_2}{\epsilon_1+\epsilon_2}(- Q_1 Q_2) \kappa  | x_1-  x_2|,  ~~~ \kappa=\frac{g_\2d^2}{4\pi},~~
\label{4p1}
\end{eqnarray}
where $Q_1$ and $Q_2$ are the charge numbers and $x_1$ and $x_2$ are
the spatial coordinates of the interacting particles.  We shall use
$\Phi_{12}$ to study the stability of the QED neutron with quarks
interacting in QED.

\vspace*{-0.2cm}
\section{ The Stability of the QED Neutron}

The success of the open-string description of the QCD and QED mesons
leads to the search for other neutral quark systems stabilized by the
QED interaction between the constituents in the color-singlet
subgroup, with the color-octet QCD gauge interaction as a spectator
field.  Of particular interest is the QED neutron with the $d$, $u$,
and $d$ quarks \cite{Won20a}.  They form a color product group of
${\bb 3}$ $\otimes$ $ {\bb 3} $ $\otimes$ $ {\bb 3}$ = ${\bb 1} \oplus
{\bb 8} \oplus {\bb 8} \oplus {\bb {10}}$, which contains a color
singlet subgroup $\bb 1$ where the color-singlet current and the
color-singlet QED gauge field reside.  In the color-singlet
$d$-$u$-$d$ system with three different colors, the attractive QED
interaction between the $d$ and $u$ quarks may overwhelm the repulsive
QED interaction between the two $d$ quarks to stabilize the composite
color-singlet QED neutron into a linear configuration.  It is
reasonable to treat the system in 1+1D space-time.  We generalize the
two-body equations of (\ref{3p1a}) and (\ref{3p1b}) to the three-quark
system by imposing three mass-shell constraints relating the momenta,
the masses, and their interactions in the form
\begin{subequations}
\label{6p3}
\begin{eqnarray}
\hspace*{-0.7cm}{\cal H}_1|\Psi\rangle =\bigl \{  p_1^2 - m_1^2 - [ \Phi_{12} (x_{12}) +\Phi_{13} (x_{13}) ]\bigr \}\,|\Psi\rangle=0,~~
\label{6p3a}
\\
\hspace*{-0.7cm}{\cal H}_2|\Psi\rangle= \bigl \{ p_2^2 - m_2^2 - [ \Phi_{21} (x_{21}) +\Phi_{23} (x_{23}) ] \bigr \} \,|\Psi\rangle=0,~~
\label{6p3b}
\\
\hspace*{-0.7cm}{\cal H}_3|\Psi\rangle= \bigl \{ p_3^2 - m_3^2 - [ \Phi_{31} (x_{31}) +\Phi_{32} (x_{32}) ]\bigr \} \,|\Psi\rangle=0,~~
\label{6p3c}
\end{eqnarray}
\end{subequations}
where $\Phi_{ij}$ is the effective two-body QED interaction
(\ref{4p1}) extracted from Schwinger's exact QED solution in 1+1D.  We
search for an energy minimum for the QED neutron state by using a
variational wave function.  It is convenient to choose a Gaussian
variational wave function of the spatial dimensionless variables $y_1,
y_2, y_3$ with standard deviations $\sigma_1$, $\sigma_2$, and
$\sigma_3$ as variational parameters, for the $d$, $u$, and $d$ quarks
respectively,
\begin{eqnarray}
\Psi(y_1,y_2,y_3)= N\exp \biggl \{ -\frac{y_1^2}{4\sigma_1^2}
-\frac{y_2^2}{4\sigma_2^2}-\frac{y_3^2}{4\sigma_3^2} \biggr \},   
\end{eqnarray}
where $y_i$=$\sqrt{\kappa}x_i$.  The charge numbers of the quarks are
$Q_1=Q_3=-1/3$, and $Q_2 =2/3$.  For the
lowest-energy state, $ \sigma_1=\sigma_3$, and  the variational
parameters consist only of $\sigma_1$ and $\sigma_2$.  We look for the
state with the lowest composite mass
$M$=$\epsilon_1+\epsilon_2+\epsilon_3$ in the variations of $\sigma_1$
and $\sigma_2$,
\begin{eqnarray}
\frac{\delta^2 M(\sigma_1,\sigma_2)}{\delta \sigma_1 \delta \sigma_2}
= 0.
\label{5p7}
\end{eqnarray}
The motion of the three quarks should maintain a fixed center of mass
for the composite system.  The coordinates of the
three quarks must satisfy the center-of-mass condition on the spatial
coordinates, $ \sum_{i=1}^3 \epsilon_i y_i=0.  $ The variational wave
function $\Psi$ is normalized according to
\vspace*{-0.2cm}
\begin{eqnarray}
\hspace*{-0.7cm}\int \!\! dy_1 dy_2 dy_3 |\Psi(y_1,y_2,y_3)|^2 \delta (\epsilon_1 y_1+\epsilon_2 y_2+ \epsilon_3 y_3) \!=\! 1.~
\end{eqnarray}

\vspace*{-0.7cm}
\begin{figure} [h]
\centering
\includegraphics[scale=0.23]{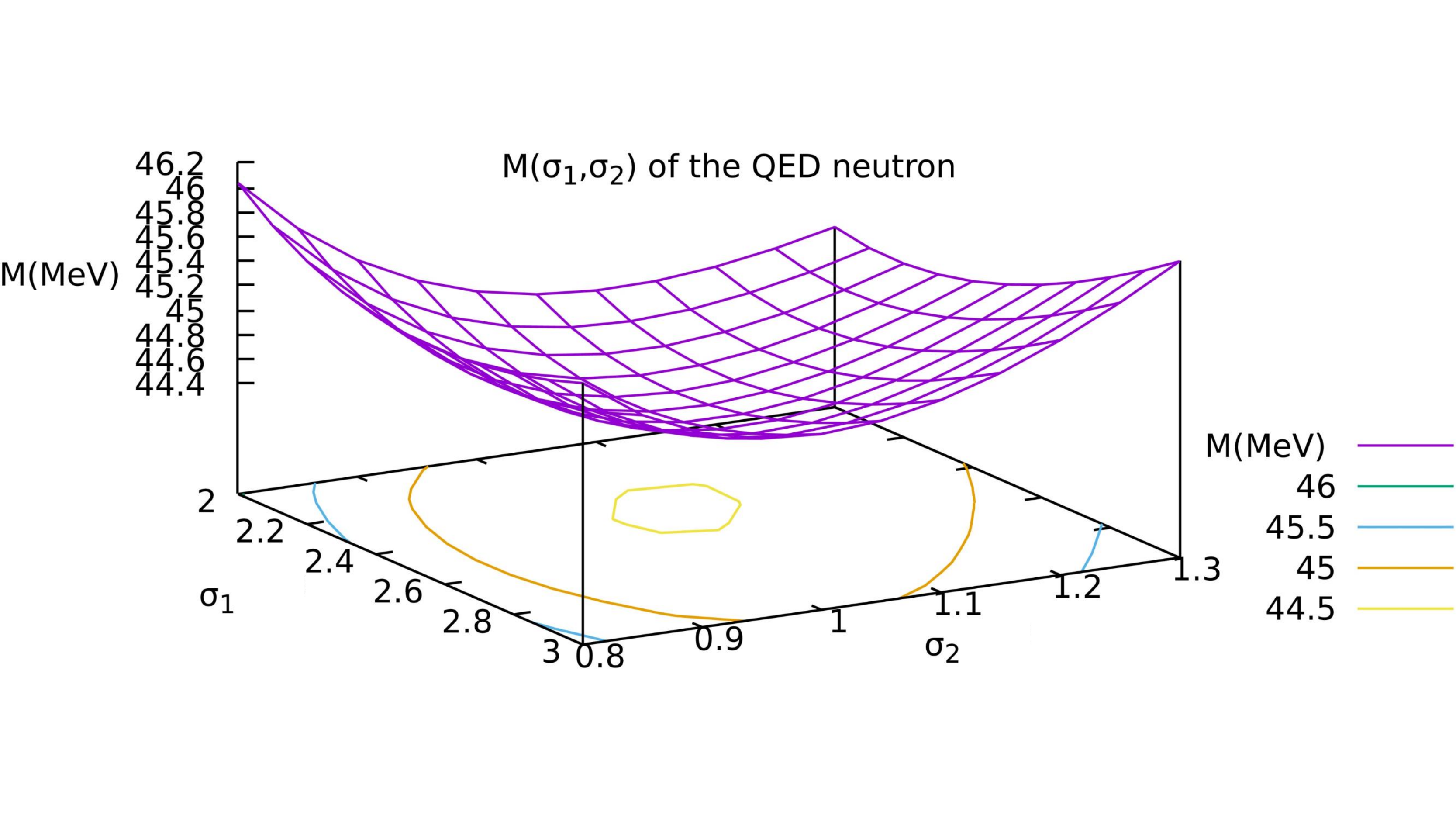}

\vspace*{-0.8cm}
\caption{The mass $M$ of the QED neutron as a function of the
  variation parameters $\sigma_1,$ and $\sigma_2$.  The QED neutron has an energy
  minimum at $M=44.5$ MeV. }
\label{figmin}
\end{figure}

\vspace*{-0.4cm} By the variational calculations, we find that the QED
neutron is stable \cite{Won20a}.  Its mass $M$ has an energy minimum, $M$=$44.5$
MeV, at $\sigma_1$=2.40 and $\sigma_2$=1.05   as shown in
Fig.\ \ref{figmin},
corresponding to  $\sigma_1/\sqrt{\kappa}$=19.9 fm and
$\sigma_2/\sqrt{\kappa}$=8.71 fm.

We study next the stability of the QED color-singlet proton with two
$u$ quarks and a $d$ quark by carrying out variational calculations
similar to those for the QED neutron.  The variation over a very large
range of $\sigma_1$ and $\sigma_2$ fails to find an energy minimum.
Extending the range of $\sigma$ will only drive the total energy of
the system lower with the $u$ quarks farther and farther apart without
the energy turning to a minimum.  An examination of the potential
energies of the interaction indicates that the QED proton is unstable
because of the stronger repulsion between the two $u$ quarks in
comparison with the weaker attractive interactions between the $d$ and
the $u$ quarks.  The QED proton also does not possess a continuum
state with isolated quarks because the isolation of quarks is
forbidden.  Therefore, the QED proton does not exist as a stable bound
state or a continuum state.

\vspace*{-0.1cm}
\section{Stability of QED Neutron against Weak Decay and Dark matter }

\vspace*{-0.1cm}

The absence of a QED proton state has an important consequence for the
weak decay of the QED neutron, which could occur when a $d$ quark
decays into a $u$ quark.  Such a QED neutron weak decay would result
in a possible QED proton final state, if a QED proton state could
exist.  Because there is no final bound or continuum QED proton state
for the QED neutron to decay onto, the density of final states for the
weak decay is zero.  Consequently the rate for the QED neutron to
decay into a QED proton is zero.  The QED neutron can only decay by a
baryon-number non-conserving transition which presumably has a very
long life time.  Therefore, the lowest energy QED neutron is a stable
particle with a very long lifetime and is in fact a dark neutron.
Because of its long lifetimes, the QED dark neutron may be good candidate
 for a part of the dark matter.
Self-gravitating
assemblies of QED dark neutrons 
may be
produced by coalescence during the deconfinement-to-confinement phase transition of
the quark gluon plasma in the evolution of the early Universe.

In  other astrophysical frontiers,
the merging of two neutron stars 
may lead to the production of a
quark matter with deconfined quarks \cite{Bau19}.
The coalescence of deconfined quarks  during
the
deconfinement-to-confinement phase transition
will produce QED neutrons in the
post-merger environment.  
It has also been suggested that deconfined
quark matter may be present in the core of a massive neutron star
\cite{Ann20}.  In such a neutron star, the transition region close to
the core may contain QED neutron matter arising from the coalescence
of deconfined quarks. 

\vspace*{-0.2cm}

\section{Conclusion and Discussions}

In a system of quarks interacting in QCD and QED interactions, the
current and the gauge fields reside in color-singlet QED and
color-octet QCD subspaces and execute independent collective
excitations.  The collective excitations in the color-octet sector
give rise to the QCD mesons such as $\pi$, $\eta$, and $\eta'$
particles. They can be described as open-string states.  In a similar
manner, collective excitations of the color-singlet sector can give
rise to QED mesons.  The energies of the lowest QED meson states have
been estimated. There are encouraging pieces of evidence for the
occurrence of QED mesons as the anomalous X17 and E38 particles
observed recently.  On-going experiments to confirm these QED mesons
are continuing.

The stability of the QED meson states leads to the study of the related
QED neutron consisting of two $d$ quarks and a $u$ quark in QED
interactions.  The attractive QED interaction between the $u$ quark
and the two $d$ quarks overwhelms the repulsion between the two $d$
quarks to stabilize the QED neutron at an estimated mass of 44.5 MeV.
The analogous QED proton has been found to be unstable, and it does
not provide a bound state for the QED neutron to decay onto by way of
the weak interaction.  Hence the QED neutron may be  stable against the
weak interaction.  It may have a very long lifetime and may be a good
candidate for the dark matter.
Because QED mesons and QED neutrons may arise from the coalescence of
deconfined quarks during the deconfinement-to-confinement phrase
transition in different environments such as in high-energy heavy-ion collisions,
neutron-star mergers  \cite{Bau19}, and neutron star cores \cite{Ann20}, the search  of the QED bound states
in various environments  
 will be of great interest.

\end{document}